\documentclass[twocolumn,pre,amsmath,amssymb]{revtex4-1}

\usepackage{graphicx}
\usepackage{dcolumn}
\usepackage{bm}
\usepackage{float}
\usepackage{mathrsfs}

\usepackage{hyperref}
\hypersetup{
 colorlinks = true,
 linkcolor = {blue},
 citecolor = {blue},
 urlcolor = {blue},
}
 
\newcommand{\cD}{\ensuremath{\mathcal{D}}}

\newcommand{\cH}{\ensuremath{\mathcal{H}}}

\newcommand{\cN}{\ensuremath{\mathcal{N}}}
\newcommand{\cP}{\ensuremath{\mathcal{P}}}

\newcommand{\cW}{\ensuremath{\mathcal{W}}}

\newcommand{\cU}{\ensuremath{\mathcal{U}}}

\newcommand{\bx}{\ensuremath{\boldsymbol{x}}}

\newcommand{\bp}{\ensuremath{\boldsymbol{p}}}

\newcommand{\br}{\ensuremath{\boldsymbol{r}}}

\newcommand{\bv}{\ensuremath{\boldsymbol{v}}}
\newcommand{\bs}{\ensuremath{\boldsymbol{s}}}

\newcommand{\bD}{\ensuremath{\boldsymbol{D}}}

\newcommand{\bI}{\ensuremath{\boldsymbol{I}}}

\newcommand{\bOm}{\ensuremath{\boldsymbol{\Omega}}}

\newcommand{\bbR}{\ensuremath{\mathbb{R}}}
\newcommand{\bbP}{\ensuremath{\mathbb{P}}}

\newcommand{\vrip}{\ensuremath{V_\textup{R}}}
\DeclareMathOperator{\kpf}{\kappa_\textup{F}}
\DeclareMathOperator{\tkpf}{\tilde{\kappa}_\textup{F}}

\begin{document}
\title{Evaluating the phase dynamics of coupled oscillators \\via time-variant topological features}
\author{Kazuha Itabashi}
\email{itabashi@biom.t.u-tokyo.ac.jp}
\affiliation{
	Graduate School of Information Science and Technology,\\
	The University of Tokyo, Tokyo 113-8656, Japan
}

\author{Quoc Hoan Tran}
\email{tran\_qh@ai.u-tokyo.ac.jp}
\affiliation{
	Graduate School of Information Science and Technology,\\
	The University of Tokyo, Tokyo 113-8656, Japan
}

\author{Yoshihiko Hasegawa}
\email{hasegawa@biom.t.u-tokyo.ac.jp}
\affiliation{
 Graduate School of Information Science and Technology,\\
	The University of Tokyo, Tokyo 113-8656, Japan
}

\date{\today}

\begin{abstract}
By characterizing the phase dynamics in coupled oscillators, we gain insights into the fundamental phenomena of complex systems.
The collective dynamics in oscillatory systems are often described by order parameters, which are insufficient for identifying more specific behaviors. 
To improve this situation, we propose a topological approach that constructs the quantitative features describing the phase evolution of oscillators.
Here, the phase data are mapped into a high-dimensional space at each time, and the topological features describing the shape of the data are subsequently extracted from the mapped points.
These features are extended to time-variant topological features by adding the evolution time as an extra dimension in the topological feature space.
The time-variant features provide crucial insights into the evolution of phase dynamics.
Combining these features with the kernel method, we characterize the multi-clustered synchronized dynamics during the early evolution stages.
Finally, we demonstrate that our method can qualitatively explain chimera states.
The experimental results confirmed the superiority of our method over those based on order parameters, especially when the available data are limited to the early-stage dynamics. 
\end{abstract}

\maketitle

\section{Introduction\label{sec:introduction}}

Coupled phase oscillators are widely used for investigating cooperative behaviors in complex systems. 
The coupling schemes of natural oscillators are reflected in the dynamic behaviors of complex systems, which include synchronization, multistability, chaos, and chimera states. 
For example, when a synchronized state interchanges with an asynchronized state during the one-day cycle of a circadian clock system, a metabolic disorder, cataplexy, or narcolepsy may be present \cite{ferri2005nrem, dibner2010mammalian}. 
An appropriate understanding of oscillator dynamics would also promote the realization of electromechanical systems, which can behave as chaotic oscillators and operate in noisy environments \cite{owens2013exactly, corron2015chaos}.
The chimera state is a counter-intuitive phenomenon in which synchronized and asynchronized states coexist in a system of identical oscillators. Chimera states have recently drawn considerable attention in studies of neural systems such as brain networks \cite{chouzouris2018chimera, cornelis2014modern,fornito2015connectomics,wang:2020:review:chimera}.
Of particular interest in coupled oscillators is predicting the attendant dynamics from the data obtained during the early stage of the system. Because many collective dynamics can be modeled by coupled oscillators, the early prediction of dynamics is potentially helpful for the diagnosis of human diseases and the detection of specific malfunctions. To realize these applications, we require  theoretical and computational methods that adequately represent the time profile of oscillators.

The intrinsic dynamic properties of coupled oscillators are often described in terms of their phase variables. Meanwhile, the degree of global synchronization is commonly expressed by the global order parameter, which represents the phase coherence of the oscillators. The global order parameter takes a value from $0$ (for complete asynchrony) to $1$ (for full synchronization)~\cite{strogatz2000kuramoto}. However, the order parameter cannot effectively analyze the synchronization situation under certain conditions. For example, when the order parameter is 0, the oscillators can be synchronized by symmetrizing their phase distribution even when they should (by definition) be completely asynchronous.

Rather than representing collective dynamics by their global parameters, our approach considers the topological aspects of phase variables, and describes the phase dynamics along the evolution timeline in terms of more efficient quantities. 
At each oscillator, we define a phase-dependent point in a high-dimensional space. We then track the time-variant evolutionary change in the shape of these points known as a \textit{point cloud} $\cP(t)$ within the space. We hypothesize that this evolutionary change is closely related to pattern formation, signal propagation, and the stochastic phenomena and extensive chaos in oscillatory systems. We thus demonstrate that tracking these changes provides crucial insights into the early-stage dynamics of the system.

Our approach presumes that topological aspects can help to reveal the underlying structure of the collective patterns of oscillators as the system evolves. Using persistent homology analysis \cite{edelsbrunner2000topological} we evaluated the shape of $\cP(t)$ in terms of its quantitative topological features and monitored the variation of these features throughout time $t$. 
Persistent homology is an algebraic topology technique that represents the shape of the data as topological structures, such as the connected components, loops, or holes in the data.
Persistent homology can effectively capture the qualitative changes in the data of various dynamical systems, including time series~\cite{tran2018topological,audun:2019:state}, time-varying networks~\cite{kim2015brain,hajij2018visual,tran:2019:scale}, and quantum data~\cite{tran:2020:topo}.

\begin{figure*}
	\includegraphics[scale=0.48]{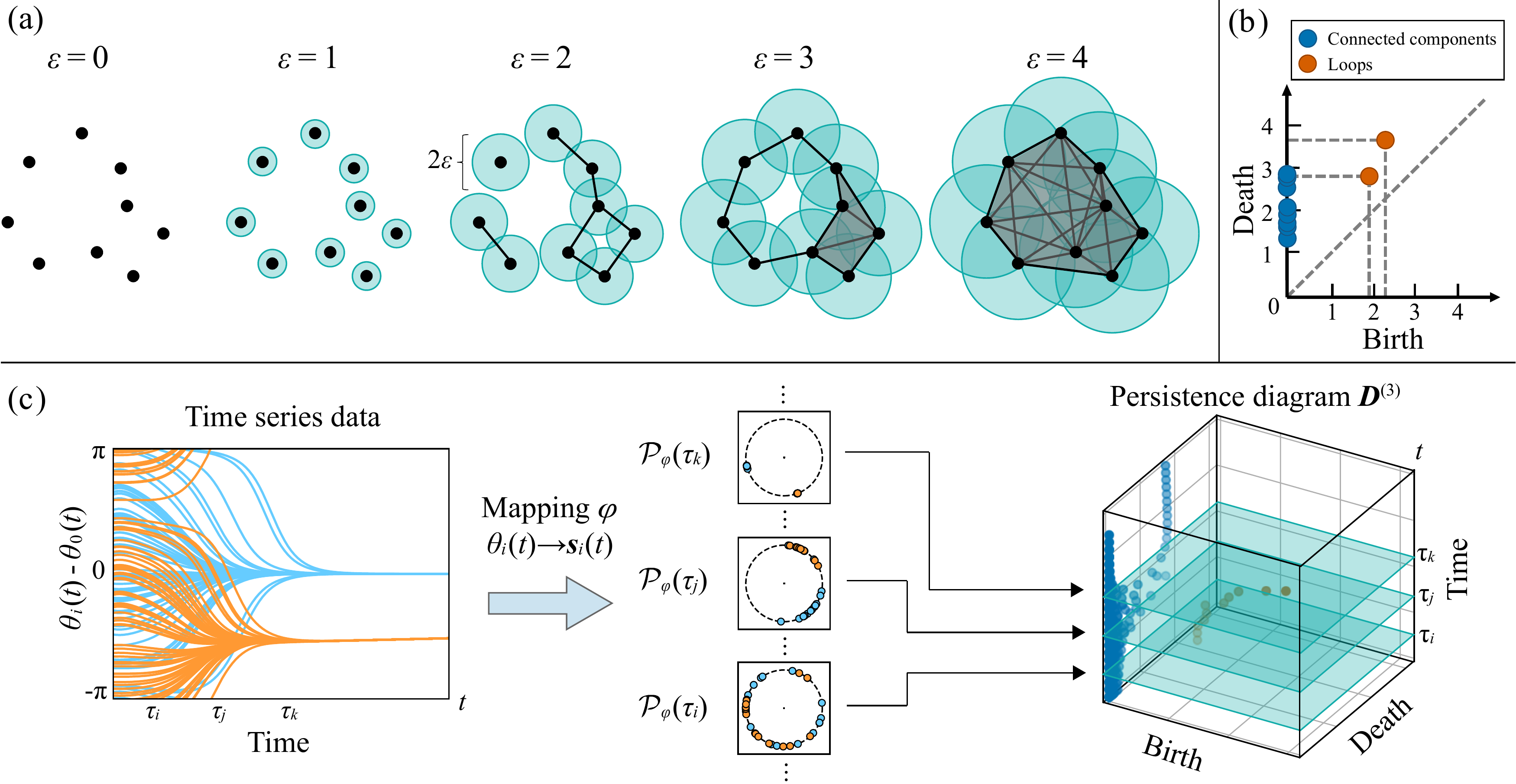}
	\caption{
Illustration of time-variant topological features for coupled oscillator systems. (a) Representative sequence of the Vietoris--Rips complex, i.e., a filtration constructed from a set of discrete points. Each ball of radius $\varepsilon$ was placed at each point, and the shape of the union of these balls was modeled with the Vietoris--Rips complex. If every pair of corresponding balls intersects, then each simplex in the complex is formed from a subset of points. The evolutionary changes in the topological structures, e.g., the merging of connected components and the emergence and disappearance of loops, were tracked by increasing $\varepsilon$ until no change was observed. At $\varepsilon=0$, nine points corresponded to nine connected components in the space. At $\varepsilon=2$, several edges were added into the complex, and a number of connected components perished and merged. The two loops that emerged at birth radii of $\varepsilon=2$ and $\varepsilon=3$ perished at death radii of $\varepsilon=3$ and $\varepsilon=4$, respectively. (b) The collection of birth and death radii is represented in a two-dimensional \textit{persistence diagram}.
(c) At each time step, the coupled oscillator phases were mapped onto a set of discrete points, i.e., a point cloud within a metric space. The two-dimensional persistence diagram was constructed from the point cloud at each time step. The time-variant topological features were obtained by concatenating the diagrams across all time steps into a three-dimensional persistence diagram.}\label{fig:tda_explain}
\end{figure*}

Given a nonnegative threshold $\varepsilon$, we place one $\varepsilon$-radius ball centered at each point in $\cP(t)$ and observe the shape of the space overlapped by these balls~[Fig.~\ref{fig:tda_explain}(a)]. When $\varepsilon$ is sufficiently small, the shape is obtained without altering the original points. As $\varepsilon$ increases, the balls intersect and the topological structures (e.g., connected components and loops) change within the space. The connected components tend to merge, while loops emerge and then vanish with gradual changes in the threshold. At each time point, we define the topological features as the values of $\varepsilon$ that represent the emergence and disappearance of the topological structures~[Fig.~\ref{fig:tda_explain}(b)]. We then extend these features by adding the time axis~[Fig.~\ref{fig:tda_explain}(c)].

These extended features, referred to as time-variant topological features, can reflect the temporal behavior of the oscillators and therefore provide useful knowledge for predicting the dynamics. In fact, these features can serve as discriminate features in qualitative evaluations of the phase dynamics of oscillators. We can input these features into machine learning kernel algorithms for statistical learning tasks such as classifying the behaviors of multicluster synchronization or predicting chimera states with co-existing synchronous and asynchronous domains. Interestingly, our approach can characterize the phase dynamics of oscillators at a very early evolutionary stage, where conventional order parameters are ineffective.

In the following sections, we introduce the time-variant topological features and the kernel method that incorporates these features into statistical learning tasks.
Using these features, we then explore dynamic behaviors such as multicluster synchronization and chimera states during the time evolution of the Kuramoto model~\cite{kuramoto:1975:oscillators,kuramoto2003chemical} for a system of oscillators.
Finally, we summarize our results and discuss interesting directions for future work.

\section{Methods\label{sec:method2}}

\subsection{Time-variant topological features}

We adopted the Kuramoto model~\cite{kuramoto:1975:oscillators,kuramoto2003chemical}, the most common and best-suited model
for understanding synchronization phenomena in physical, chemical, and biological systems. The Kuramoto model formulates a set of $N$ coupled heterogeneous oscillators as a set of first-order differential equations:
\begin{align}\label{eqn:kura}
 \dfrac{d\theta_i}{dt} = \omega_i + \frac{\kappa}{N}\sum_{j=1}^{N}g_{ij}\sin(\theta_j-\theta_i-\alpha).
\end{align}
Here, $\theta_i$ and $\omega_i$ denote the phase and natural frequency of oscillator $i$, respectively, and $g_{ij}\geq 0$ represents the coupling strength between oscillators $i$ and $j$. The parameter $\kappa$ is the global coupling constant and the angle $\alpha$ is a tunable parameter describing the phase lag. 

To study the dynamics of the coupled oscillators, we constructed a mapping $\varphi$ from the set $\{\theta_1(t),\theta_2(t),\ldots, \theta_N(t)\}$ of phases to the set $\cP_{\varphi}(t)=\{\bs_1(t),\bs_2(t),\ldots,\bs_N(t)\}$ of points in an $L$-dimensional space $\mathbb{R}^L$, where 
$\bs_i(t)=\varphi(\theta_i(t)) \in \mathbb{R}^L$. 
For an appropriate mapping $\varphi$, the information on the evolutionary change in the configuration of $\cP_{\varphi}(t)$ provides important insights into the system dynamics. 
To quantify the features in the configuration of $\cP_{\varphi}(t)$, we applied persistent homology theory.
We defined a distance function $d_{\varphi}: \cP_{\varphi}(t) \times \cP_{\varphi}(t) \rightarrow \bbR$ to evaluate the dissimilarity between oscillators $i$ and $j$ at time $t$. 
We then centered a $\varepsilon$-radius ball at each point $\bs_i(t)$ in $\cP_{\varphi}(t)$, i.e., 
to form the set $B(\varepsilon, \bs_i(t))=\{\bv \in \bbR^L \mid d_{\varphi}(\bs_i(t), \bv) \leq \varepsilon\}$. 
Taking the union of these balls, we obtain the overlapped space
\begin{align}\label{eqn:ball}
\cU(\varepsilon, \cP(t))=\bigcup_{i=1}^N B(\varepsilon, \bs_i(t)).
\end{align}
The shape of this space represents the configuration of $\cP_{\varphi}(t)$ at radius $\varepsilon$. Persistent homology tracks the changes in this shape as the radius $\varepsilon$ increases.

Methods in computational topology allow for modeling the shape of $\displaystyle{\cU(\varepsilon, \cP_{\varphi}(t))}$ in a mathematically and computationally tractable representation, i.e., a simplicial complex, defined as a complex of geometric structures called simplices. 
Here, an $n$-simplex represents a generalization of the notion of a triangle or tetrahedron to arbitrary dimensions. 
The convex hull of any nonempty subset of the vertices defining a simplex is called a face of the simplex.
For example, a $0$-simplex is a point, a $1$-simplex is a line segment (with faces comprising both end points), and a $2$-simplex is a triangle and its enclosed area (with faces composed of three edges and three vertices). Similarly, a $3$-simplex is a filled tetrahedron (with faces composed of triangles, edges, and vertices), and a $4$-simplex (beyond visualization) is a filled shape with faces composed of tetrahedrons, triangles, edges, and vertices.

A main type of simplicial complex is the Vietoris--Rips complex, which we now briefly review.
A Vietoris--Rips complex $\vrip(\varepsilon, \cP_{\varphi}(t))$ is a collection of simplices,
where each simplex is built over a subset of points in $\cP_{\varphi}(t)$ 
provided that $B(\varepsilon, \bs_i(t))\cap B(\varepsilon, \bs_j(t)) \neq \varnothing$ for every pair of points $\bs_i(t), \bs_j(t)$ in the subset.
When $\varepsilon=0$, the complex contains only $0$-simplices, i.e., the discrete points. As $\varepsilon$ increases, the points become connected and edges ($1$-simplices) and filled triangles ($2$-simplices) are introduced into the complexes. 
At some considerably large $\varepsilon$, all points become interconnected and no useful information is conveyed.
The sequence of embedded complexes obtained by this process is called a filtration.  

Persistent homology focuses on the emergence and disappearance of topological structures such as connected components and loops in the filtration. For example, in Fig.~\ref{fig:tda_explain}(a), there were nine connected components in the space at $\varepsilon=0$ and $\varepsilon=1$. Several components were merged at $\varepsilon=2$, meaning that six connected components were destroyed and three connected components remained in the space. Similarly, one loop appeared at $\varepsilon=2$, which disappeared at $\varepsilon=3$. Another loop that appeared at $\varepsilon=3$ disappeared at $\varepsilon=4$.
These topological structures are mathematically represented by $0$- and $1$-dimensional persistent homology groups, which are vector spaces with dimensions corresponding to the number of connected components and the number of loops, respectively~\cite{edels:2010:topobook}. 
Using the emergence and disappearance of topological structures in the filtration, we quantified the evolving shape of $\displaystyle{\cU(\varepsilon, \cP_{\varphi}(t))}$. Each topological structure was assigned to a persistent pair of radii $(\varepsilon_b, \varepsilon_d)$ that originated at birth radius $\varepsilon=\varepsilon_b$ and perished at death radius $\varepsilon=\varepsilon_d$. The collection of all persistent pairs in a two-dimensional coordinate was presented as a two-dimensional persistence diagram $\bD^{(2)}(\cP_{\varphi}(t))$ containing the topological features of $\cP_{\varphi}(t)$ [Fig.~\ref{fig:tda_explain}(b)]. We proposed the time-variant topological features containing time-related information on the dynamics, along with the birth and death radii. 
We constructed three-dimensional persistence diagrams by concatenating the two-dimensional persistence diagrams along the time-axis at time steps $\tau_0 < \tau_1 < \ldots < \tau_{T-1}$ [Fig.~\ref{fig:tda_explain}(c)]. Mathematically, this construction is described by 
\begin{align}
 \bD^{(3)}(\varphi) = \{(\varepsilon_b, \varepsilon_d, \tau) \mid (\varepsilon_b, \varepsilon_d) \in \bD^{(2)}(\cP_{\varphi}(\tau)),\nonumber\\ \tau=\tau_0, \tau_1, \ldots, \tau_{T-1}\}.
\end{align}

\subsection{The kernel method for topological features\label{sec:kernel}}

In statistical-learning tasks using time-variant topological features,
we typically quantify the underlying patterns in a collection of inputs $\cD = \{\bD_1, \ldots, \bD_M\}$ from a certain set of diagrams, and use these patterns to evaluate previously unseen data. As a persistence diagram is a multiset of points of variable size, algorithms that take vector inputs or that require the inner product of the data are not easily applicable. Instead, we employed kernel methods that take the similarity measure $\kappa(\bD_i, \bD_j)$ between any two diagrams $\bD_i, \bD_j$. More precisely, a function $\kappa: \cD \times \cD \to \bbR$ is called a kernel if the kernel Gram matrix $K$ with entries $K_{ij}=\kappa(\bD_i, \bD_j)$ is positive semidefinite. To define a kernel, a feature map $\Phi$ was constructed by mapping a diagram $\bD_i$ to a vector $\Phi(\bD_i)$ in a Hilbert space $\cH_b$, in which we can define the inner product $\langle\cdot,\cdot \rangle_{\cH_b}$. 
Every feature map $\Phi$ defines the kernel $\kappa(\bD_i, \bD_j) := \langle\Phi(\bD_i),\Phi(\bD_j) \rangle_{\cH_b}$. 

As the kernel for two-dimensional persistence diagrams, researchers have proposed the persistence-scale space kernel~\cite{reininghaus:2015:mskernel} based on the heat diffusion kernel, the persistence-weighted Gaussian kernel~\cite{kusano:2016:weightgauss}, which emerged from kernel mean embedding, the sliced Wasserstein kernel formulated in Wasserstein geometry~\cite{mathieu:2017:slicekernel}, and the persistence Fisher kernel~\cite{le:2018:pfk}, which relies on Fisher information geometry.
From theoretical and practical perspectives, the persistence Fisher kernel is a superior choice for distinguishing persistence diagrams. For instance, it computes over the number of points in the diagram with linear time complexity.
It also outperforms the other kernel methods on various benchmarks~\cite{le:2018:pfk}.
We extended the persistence Fisher kernel to three-dimensional persistence diagrams, as briefly explained below.

The persistence Fisher kernel considers each persistence diagram as the sum of normal distributions and then measures the similarity between the distributions via the Fisher information metric. 
Each persistence diagram $\bD$ can be considered as a discrete probability mass $\mu_{\bD}=\sum_{\bp \in \bD}\delta_{\bp}$, where $\delta_{\bp}$ is the Dirac measure centered on $\bp$.
We can smooth and normalize $\mu_{\bD}$ by the summation $\rho_{\bD}$ of normal distributions as
\begin{align}\label{eqn:rho}
    \rho_{\bD}=\sum_{\bp \in \bD}\dfrac{1}{Z}\cN(\bp, \nu\bI),
\end{align}
where $\cN(\bp, \nu\bI)$ is the Gaussian function centered at $\bp$ with bandwidth $\nu$, $\bI$ is an identity matrix, and $Z=\int_{\bOm}\sum_{\bp \in \bD}\cN(\bx;\bp, \nu\bI)d\bx$ is the normalization constant with the integral calculated on a domain $\bOm$.
Given a positive $\beta$, we define the following kernel
\begin{equation}\label{eqn:kpf}
 \tkpf(\bD_i, \bD_j) = \text{exp}(-\beta d_{\textup{F}}(\bD_i, \bD_j)),
\end{equation}
where 
$
 d_{\textup{F}}(\bD_i, \bD_j) = \text{arccos}\left( \int_{\bOm} \sqrt{\rho_{\bD_i}(\bx)\rho_{\bD_j}(\bx)}d\bx\right)
$
is the Fisher information metric between $\rho_{\bD_i}$ and $\rho_{\bD_j}$ (see Appendix \ref{appx:kernel}).
We set $\beta=1.0$ in our experiments.

The kernel $\tkpf(\bD_i, \bD_j)$ takes a value in $(0, 1]$ and equals 1 if $\bD_i$ and $\bD_j$ are the same.
However, this definition is ill-defined if one diagram is empty, and must be modified 
to deal with such cases.
Suppose that $\bD_j$ is empty and $\bD_i$ contains only one element $\bp=(b_1, d_1, \tau_1)$. The kernel should approximate 1 if $d_1 - b_1$ approximates zero. Now consider $\rho_{\bD^\prime_j}$, where $\bD^\prime_j$ is the set of elements $\bp^\prime=\left(\frac{b_1+d_1}{2}, \frac{b_1+d_1}{2}, \tau_1\right)$. Each $\bp^\prime$ is the projected point of $\bp$ on the diagonal plane $\cW=\{a, a, \tau \mid a, \tau \in \bbR\}$.
Let $\bD_{i\Delta}$ and $\bD_{j\Delta}$ be the point sets obtained by projecting two persistence diagrams $\bD_i$ and $\bD_j$ on $\cW$. The
kernel compares two extended diagrams, 
$\bD'_i=\bD_i\cup \bD_{j\Delta}$ and $\bD'_j=\bD_j\cup \bD_{i\Delta}$,
containing the same number of points.
Therefore, we can write $\bOm=\bD_i\cup \bD_{i\Delta}\cup\bD_j\cup \bD_{j\Delta}$, and the persistence Fisher kernel becomes 
\begin{align}\label{eqn:kpf:extend}
    \kpf(\bD_i, \bD_j) = \tkpf(\bD^\prime_i, \bD^\prime_j) = \text{exp}(-\beta d_{\textup{F}}(\bD^\prime_i, \bD^\prime_j)).
\end{align}

Under this kernel, persistence diagrams are considered close if the points that are distant from the diagonal plane in the two diagrams belong to very near regions in the space. Conversely, these diagrams are significantly different if the points distant from the diagonal plane exhibit two significantly different distributions in the two diagrams.
We can perform kernel principal component analysis (kPCA), an extension of principal component analysis (PCA) that uses kernel methods~\cite{shawe2005eigenspectrum}.
As our data $\bD_1, \ldots, \bD_M$ are mapped into the feature space as $\Phi(\bD_1), \ldots, \Phi(\bD_M)$, we can perform PCA on the covariance matrix of the mapped data. Even when the explicit form of $\Phi$ is unknown, the method in kPCA reduces to finding the eigenvalues and eigenvectors of the kernel Gram matrix.

\section{Results\label{sec:results}}

\subsection{Multicluster synchronization\label{subsec:resultA}}
A network system of multiple coupled oscillators can demonstrate multicluster synchronization, i.e., the network may split into several clusters of independent synchronized or organized behavior rather than form an entire system of synchronized behavior. Multicluster synchronizations are found in asymptotic states~\cite{jalan:2003:prl:organized, jalan:2005:pre:synchronized, amritkar:2005:pre:synch}, in transient states~\cite{arenas:2006:prl:synch}, and in modular and hierarchical structures~\cite{zhou:2006:prl:hieararchical}. Here, we demonstrate that our time-variant topological features obtained at the early stage of the dynamics can help in predicting the multicluster synchronized behavior of oscillators.

\begin{figure}
\centering
	\includegraphics[width=8.5cm]{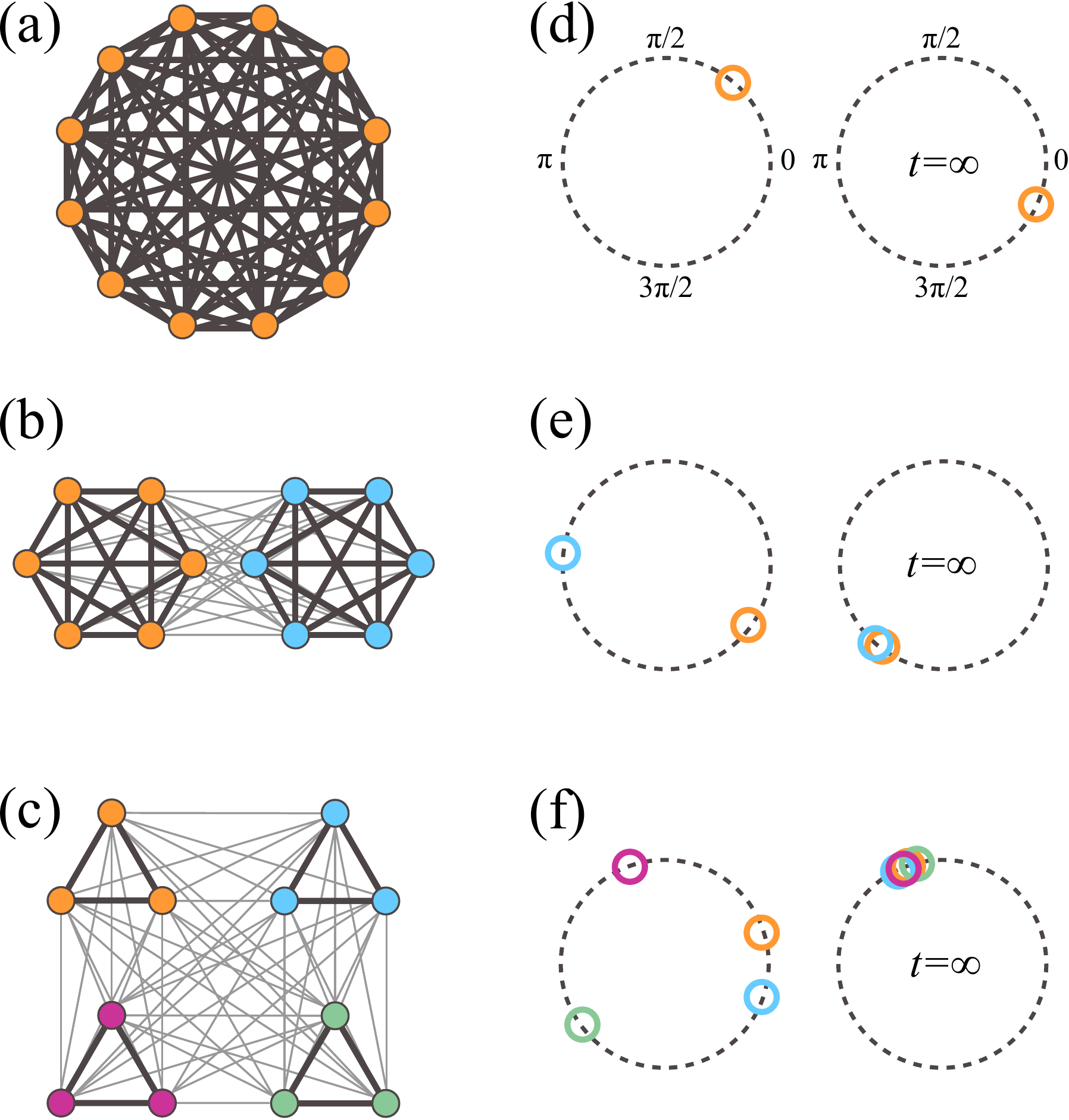}
\caption{
Schematic of oscillator networks with different coupling configurations and their corresponding synchronization behaviors in the middle and final (infinite time) stages. The vertices and edges in (a), (b), and (c) represent the oscillators and their coupling relations, respectively. Bold and thin edges imply strong coupling ($g_{ij}=2$) and weak coupling ($g_{ij}=0.01$), respectively. (a) All oscillators were symmetrically interactive and were globally coupled with the same coupling strength. In the two-module (b) and four-module (c) networks, the coupling was stronger between oscillators in the same module than in those belonging to different modules. The oscillators exhibited different synchronized behaviors at sufficiently large times in the mid-evolution stage: (d) single-cluster, (e) two-cluster, and (f) four-cluster synchronization.}
\label{fig:modelA}
\end{figure}

We configured oscillator networks with three different coupling configurations but with the same constant $\kappa=1$: 
a globally coupled network in which all coupling strengths are equal [Fig.~\ref{fig:modelA}(a)], a two-module coupled network [Fig.~\ref{fig:modelA}(b)], and a four-module coupled network [Fig.~\ref{fig:modelA}(c)]. In the globally coupled network, we set $g_{ij}=2$ for $\forall i \neq j$. In the modular coupled networks, 
we set $g_{ij}=2$ for the oscillators in the same module and $g_{ij}=0.01$ for those in different modules.
We also set the number of oscillators to $N = 128$, the angular frequency $\omega_i = 1$ for all $i$, and the tunable parameter to $\alpha=0$. In these configurations, different synchronization behaviors emerged at sufficiently large times in the middle stage of the evolution: single-cluster [Fig.~\ref{fig:modelA}(d)], two-cluster [Fig.~\ref{fig:modelA}(e)], and four cluster [Fig.~\ref{fig:modelA}(f)] synchronizations.

Equation~\eqref{eqn:kura} was numerically solved with randomly initialized phases $\theta_j(0)\in [0, 2\pi)$ and $\theta_j(t)$ was recorded at each time interval $\Delta\tau=0.8$. The time sequence $\tau_0,\tau_1,\cdots,\tau_{T-1}$ was used in the persistence diagram calculations, where $\tau_0 = 0, \tau_k-\tau_{k-1}=\Delta\tau$ $(k = 1,\ldots,T-1)$, and $T$ is the number of time steps.
Through the mapping $\varphi: \theta \to (\cos\theta, \sin\theta)$,
the set of oscillator phases was transformed to the point cloud $\cP_{\varphi}(t)=\{\bs_1(t), \bs_2(t), \ldots, \bs_N(t)\}$, where $\bs_j(t)=(\cos\theta_j(t), \sin\theta_j(t))$ lay on the unit circle in two-dimensional space. The shortest distance between $\bs_j(t)$ and $\bs_k(t)$ along the unit circle was adopted as the distance function. In our practical implementation, $\varepsilon$ in Eq.~\eqref{eqn:ball} only takes values between 0 and $\pi/2$.

\begin{figure*}
\centering
	\includegraphics[width=17.5cm]{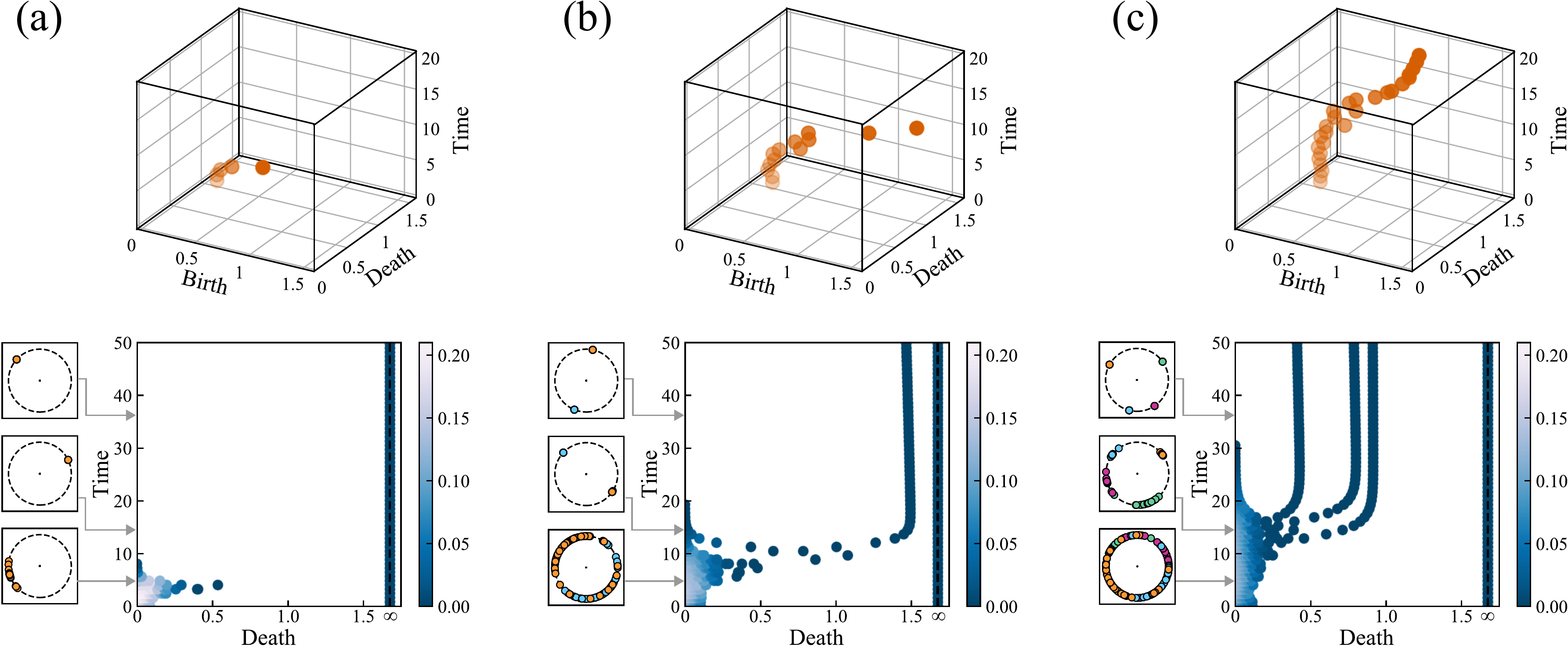}
\caption{
Examples of time-variant topological features corresponding to the coupled oscillator dynamics in Fig.~\ref{fig:modelA} (a), (b), and (c), respectively. The top row presents the three-dimensional persistence diagrams of the loop patterns, and the bottom row plots the distributions of death radii of the connected components during the phase evolution of the oscillators.
In the bottom row, the color bar at the right of each plot represents the probability densities of the death radius distribution, and the point clouds at the left show the phase plots of the coupled oscillators at $t=4.8,14.4$, and $35.2$.
When the oscillators approached a single synchronized-state cluster,
the loops quickly disappeared, and the connected components quickly merged into one component~[Fig.~\ref{fig:3dpd}(a)]. When the oscillators divided into multiclusters of synchronized states, the birth radii of the loops increased, and the more connected components survived for a longer period of evolution time~[Fig.~\ref{fig:3dpd}(b)(c)]. The number of long-lived components corresponded with the number of clusters in the synchronized state.}
\label{fig:3dpd}
\end{figure*}

Figure~\ref{fig:3dpd} shows representative time-variant topological features obtained from the globally coupled~[Fig.~\ref{fig:3dpd}(a)],
two-module coupled~[Fig.~\ref{fig:3dpd}(b)], and four-module coupled~[Fig.~\ref{fig:3dpd}(c)] networks. 
The temporal transitions in the dynamics appear as transitions in the temporal patterns of the topological features. Along the top row of Fig.~\ref{fig:3dpd}, the orange points represent the loops formed along the phase-evolution timeline. In the globally coupled network, the loops quickly disappeared as the oscillators approached a synchronized state, but in the modular coupled networks, the birth radii of the loops increased as the oscillators divided into multisynchronized, uniformly dispersed clusters.
Regarding the evolution of the connected components, the birth radius was zero because the $N$ components corresponding to $N$ oscillators appeared first. Therefore, we focused on the evolution of the death radii (bottom row of Fig.~\ref{fig:3dpd}).
As represented by the right column in Fig.~\ref{fig:3dpd} (at index $\infty$), at each time $t$, one connected component was always retained at a sufficiently large radius $\varepsilon$ in Eq.~\eqref{eqn:ball}. For instance, if $\varepsilon \geq \pi/2$, then at each time $t$, only one connected component remained in the simplicial complex created by the oscillators.

To distinguish the synchronized behavior in each coupling configuration, we examined the merging process of other connected components along the evolution timeline. In the globally coupled network, the connected components quickly merged into a single component because the oscillators reached a single synchronized-state cluster. At $t>8$, only one component existed~[Fig.~\ref{fig:3dpd}(a)]. When the oscillators were distributed as multiclusters in the synchronized state, more connected components existed over a longer time. The number of such components corresponded to the number of clusters in the synchronized state. For example, two connected components survived from $t=20$ to $t=50$~[Fig.~\ref{fig:3dpd}(b)], and four connected components survived from $t=30$ to $t=50$~[Fig.~\ref{fig:3dpd}(c)]. These values corresponded to the behaviors of the two-cluster~[Fig.~\ref{fig:modelA}(b)] and four-cluster~[Fig.~\ref{fig:modelA}(c)] synchronizations, respectively. 
In Appendix~\ref{appx:assym}, we demonstrate that time-variant topological features are also useful for investigating asymmetric networks of oscillators.

\begin{figure*}
\centering
	\includegraphics[width=17.5cm]{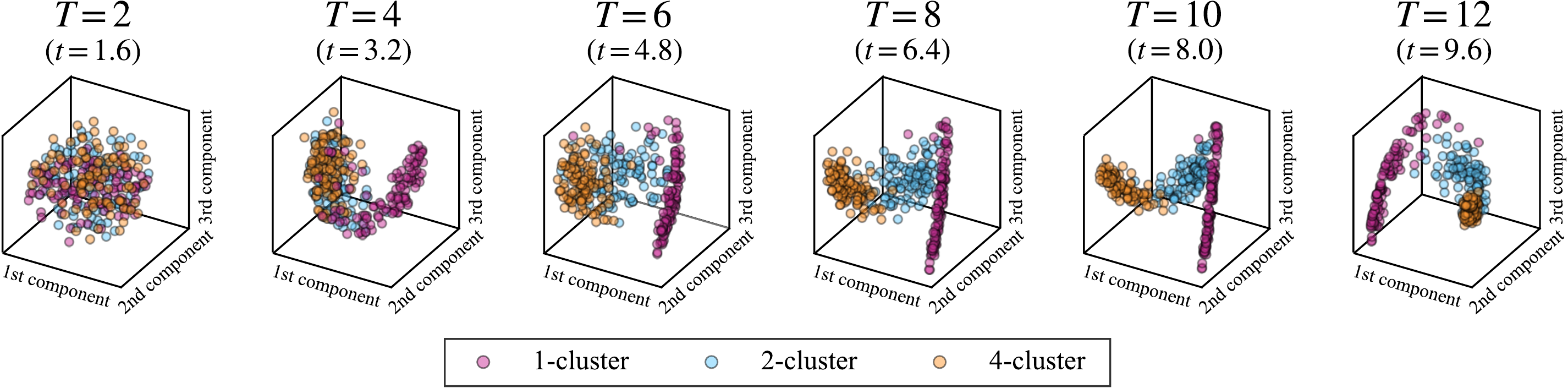}
\caption{
Projection of the kernel principal components of time-variant topological features obtained at $\tau=\tau_0, \tau_1, \ldots, \tau_{T-1}$,
where $\tau_0=0, \tau_k-\tau_{k-1}=\Delta\tau=0.8$ $(k=1,2,\ldots,T-1)$.
The number of oscillators was $N = 128$, the angular frequency was $\omega_i = 1$ for all $i$, and the tunable parameter was $\alpha=0$. Shown are the projections at $T=2,4,6,8,10$, and $12$ (left to right). The data of different synchronization schemes are shown in different colors: purple (single cluster), blue (two clusters), and orange (four clusters).
}
\label{fig:kpca_const}
\end{figure*}

Applying the kernel method, we then characterized the differences among the synchronized behaviors. Our approach does not label the synchronization behaviors a priori, but characterizes their differences in the kernel space. This approach aligns with unsupervised learning schemes, which fundamentally differ from the supervised learning schemes often used in machine learning methods. In supervised learning schemes, the learning machine is first trained on samples with predefined labels, and then attempts to predict the unknown label of a given sample. Good performance demonstrates that the learning has been generalized to samples not encountered before. In contrast, unsupervised schemes require no prior labeling but characterize the unknown dynamics via dimensional reduction methods. 
In our study, the synchronized behaviors were identified by projecting time-variant topological features onto a lower-dimensional space via kPCA of the kernel Gram matrix.

We prepared 300 coupling configurations with 100 random initializations for the oscillator phases with single, two-module, and four-module synchronizations. Single-cluster, two-cluster, and four-cluster synchronizations were observed in the time evolutions. 
The persistence diagrams were computed over a time sequence of $\tau_0,\tau_1,\cdots,\tau_{T-1}$, where $\tau_0=0$ and $\Delta\tau=0.8$. 
The parameter $T$ controls the periodicity of detecting  the synchronized behaviors. Prior to kPCA, the kernel Gram matrix of the 300 three-dimensional persistence diagrams was computed by Eq.~\eqref{eqn:kpf:extend}. 
Figure~\ref{fig:kpca_const} displays the projections up to the third principal component.  
In the early evolutionary stages ($T=2, 4$), the points belonging to different groups of synchronized behaviors were not well separated in the principal component space. However, the separability increased at later times ($T=6, 8, 10, 12$).

Accordingly, we could quantify the extent to which the time-variant topological features identified the synchronization behaviors. Here we compared this extent with those of other methods based on global order parameters, local order parameters, and dynamic connectivity. The latter method is described in Appendix~\ref{appx:classify} and Ref.~\cite{arenas:2006:prl:synch}.
The global order parameter is defined as
\begin{align}\label{eqn:global:kuramoto}
 r(t)=\frac{1}{N}\left|\sum_{j=1}^{N}e^{i\theta_j(t)}\right|,
\end{align}
where $r(t)$ takes values in the range $[0, 1]$. Note that $r(t)=1$ when all oscillators are completely synchronized. 
The local order parameter for each oscillator ~\cite{wolfrum:2011:spectral,omelchenko:2011:loss} is defined as
\begin{align}\label{eqn:kura:local}
 l_j(t)=\left|\frac{1}{2\delta+1}\sum_{|j-k| \leq\delta}e^{i\theta_k(t)}\right|,
\end{align}
where $j=1,2,\cdots,N$, and $\theta_k$ represents the phase of oscillator $k$ in a region of side length $2\delta+1$ centered at oscillator $j$. The local order parameter describes the local synchronization properties of the oscillators, and is controlled by the parameter $\delta$.

The performances of the compared methods were determined under the supervised learning scheme. 
The data were classified into three labels representing single-cluster, two-cluster, and four-cluster synchronizations. The accuracy of classifying the test data was evaluated in each model.
In the case of constant angular frequency (Fig.~\ref{fig:kpca_const}), the topological features characterized the synchronization schemes during the early-stage dynamics. We further demonstrated the effectiveness of the topological features under more flexible settings, in which the natural frequency of each oscillator followed the standard normal distribution $\cN(0,1)$.
The data in the order parameter vector, eigenvalue-spectrum vector of the dynamic connectivity matrix, and kernel Gram matrix of the persistence diagrams were classified by support vector machine (SVM) method~\cite{vapnik:1963:pattern} (see Appendix~\ref{appx:classify}). 

\begin{figure}
    \centering
    \includegraphics[width=8.5cm]{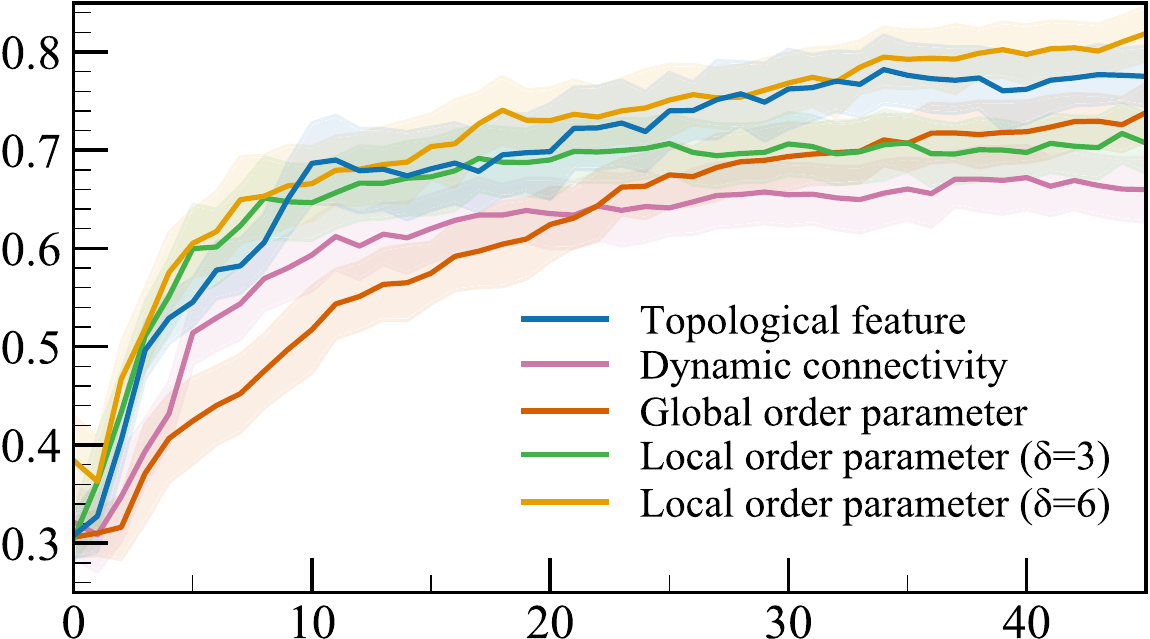}
    \caption{
Classification of synchronized states based on the time-variant topological features (blue), dynamic connectivity (purple), global order parameters (orange), and local order parameters with $\delta=3$ (green) and $\delta=6$ (yellow).
The natural frequency of each oscillator follows a standard normal distribution $\cN(0,1)$.
The lines depict the average test accuracy across 100 random train--test splits of the data.
The shaded areas indicate the confidence intervals (one standard deviation) calculated in the same ensemble of runs.}
\label{fig:gauss_acc}
\end{figure}

Figure~\ref{fig:gauss_acc} shows the average test accuracies of the methods over 100 train--test random splits at each value of $T$.  
In this task, the 100 realizations were randomly split into 50 realizations for training and 50 realizations for testing.
At $T>40$, the classification accuracy based on topological features reached nearly 76\%, versus nearly 64\% for the dynamic connectivity matrix and global order parameters. 
The classification results of the local order parameters significantly depended on the value of the tunable parameter $\delta$, which controls the region over which the local property of each oscillator is evaluated.
In the local order parameter method with $\delta=3$ and $\delta=6$, the classification accuracy was generally lower and higher than our topological feature method, respectively.
These trends occurred because the local order parameters capture the local features describing the synchronization behavior. 
Therefore, the local order parameters are effective when the connectivity and strength of the coupling depend on the oscillator locations, but $\delta$ must be appropriately adjusted.
If $\delta$ is too small, the information is restricted to the individual oscillators,
whereas if $\delta$ is too large, only the global information is available, and the method becomes equivalent to that based on the global order parameter.
In this sense, our approach provides a versatile and independent classification method that avoids the parameter-adjustment bottleneck.

In Appendices~\ref{appx:interval} and~\ref{appx:noscillators},
we investigate how varying the time interval $\Delta\tau$ and randomly selecting the number of oscillators influence the classification result. 
We confirmed that lowering the sampling interval improves the classification accuracy. We also demonstrated the impact of collecting time-variant topological features over a wide temporal range.
Overall, the topological features well described the collective behaviors even when calculated from a sparse configuration of oscillators.

\subsection{Chimera states}

\begin{figure*}
	\includegraphics[width=17cm]{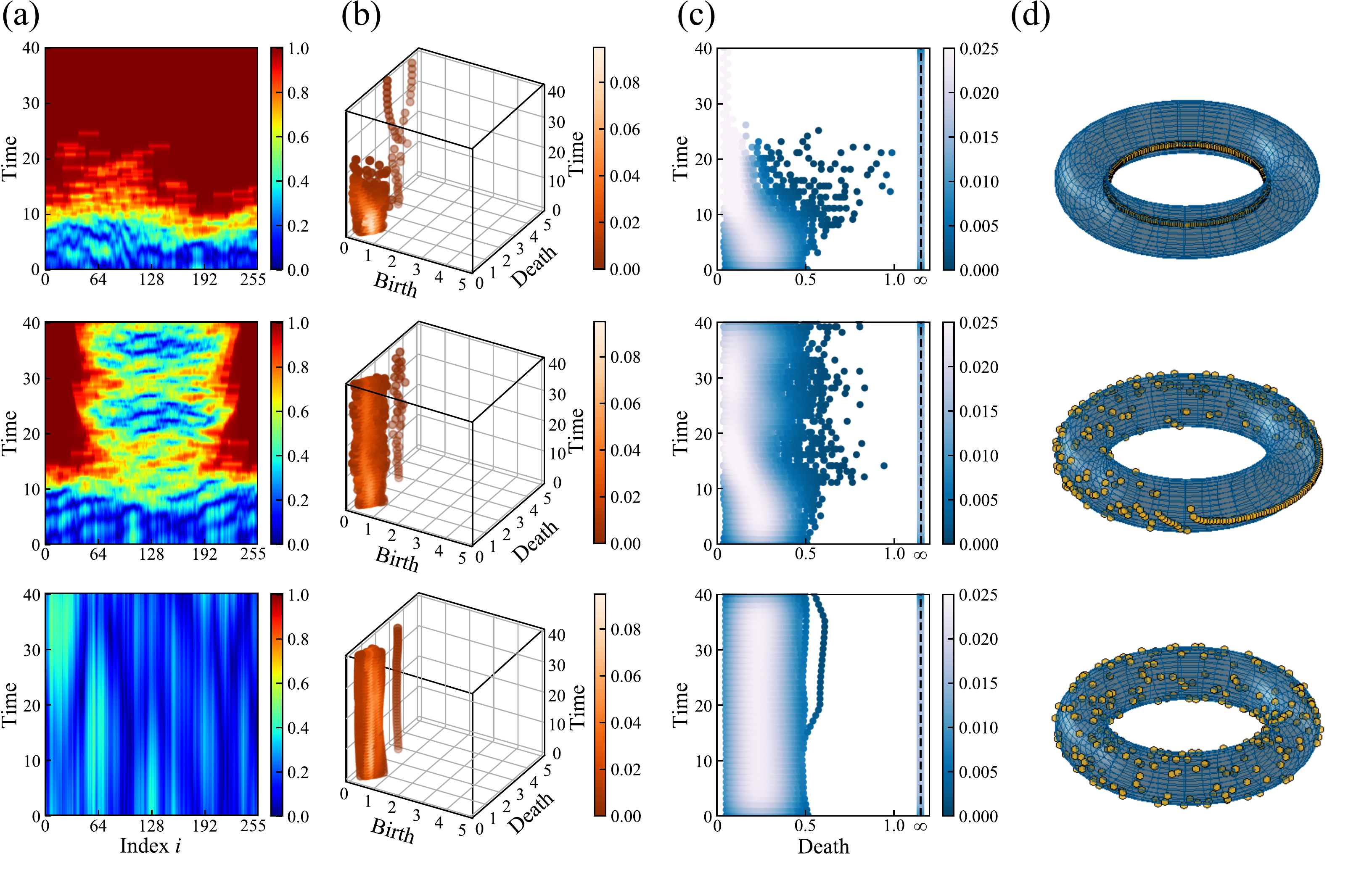}
	\caption{
	Examples of different phase dynamics along the evolution: synchronized state (top row), chimera state (middle row), and asynchronized state (bottom row). Their differences can be evaluated by tracking the time trace of local order parameters (a), or studying time-variant topological features such as the three-dimensional persistence diagrams of loop patterns (b), and the distribution of the death radii of the connected components (c). 
	The colored bar for each plot in (b)(c) represents the density of the points in the corresponding plot.
	The persistence diagrams were obtained from the mapped points of the oscillator phases on a torus surface. (d) The shape of these mapped points corresponds with each phase of dynamics at $t=40$.}
	\label{fig:lop_pd}
\end{figure*}

Using the time-variant topological features, we investigated the evolutions of the coherent and incoherent dynamics in the oscillatory systems, including the chimera states. In a chimera state, the oscillators form a region of mutually coherent populations and another region of incoherent populations~\cite{omelchenko:2008:chimera, panaggio:2015:chimera}. The chimera states were initially explored in homogeneous oscillator systems~\cite{kuramoto:2002:coexistence,daniel:2004:chimera}, and were then demonstrated experimentally~\cite{tinsley:2012:chimera,hagerstrom:2012:experimental} to establish their connection with real-world systems such as human brain networks~\cite{bansaleaau:2019:cognitive}. Referring to previous works ~\cite{zhu2014chimera,yao2015emergence}, we generated chimera states in the Kuramoto model by setting the constant $\kappa=\dfrac{\pi\eta}{\gamma}$ and the coupling strength $g_{ij}$ in Eq.~\eqref{eqn:kura} to
\begin{align}
 g_{ij} & = \begin{cases}
 \dfrac{1}{2\eta}, & {\rm if}
~\dfrac{2}{N}|i-j|>\eta\\
 0, & {\rm otherwise}
 \end{cases}.
\end{align}
Here, $\gamma$ is a tunable parameter characterizing the coupling strength among the oscillators, and $\eta\in[0,1]$ controls the range of non-local coupling. In our numerical simulation, we set $N=256$, $\omega_i=0$, $\alpha=1.39$, $\eta=0.6$, and $\gamma=0.6$ for the synchronized and chimera states, and $\gamma=6$ for the asynchronized states.

The coherence--incoherence transition was observed in the time trace of the local order parameter $l_j(t)$ for each oscillator, given by Eq.~\eqref{eqn:kura:local}.
The local order parameter quantifies the degree of the coherent and incoherent regions  around an oscillator and describes the local properties of the chimera states. More specifically, if oscillator $j$ at time $t$ exists in the coherent domain, $l_j(t)\approx 1$, but if oscillator $j$ exists in the incoherent domain, then $l_j(t)$ will be much lower. Figure~\ref{fig:lop_pd}(a) shows three time profiles of the local order parameters with $\delta=12$. In the first example, $l_j(t)$ approximated $1$ for all $j$ and $t > 10$, indicating that the oscillators became globally synchronized at $t>10$. In the second example, the $l_j(t)$ values of oscillators $80\leq j\leq 200$ were below 0.6 while the $l_j(t)$ values of the other oscillators exceeded $0.8$ at $t>10$. Based on the $l_j(t)$ values, the oscillators were roughly divided into a coherent area and an incoherent area, indicating the emergence of a chimera state. In the final example, the incoherent area dominated as all $l_j(t)$ values were lower than 0.5.

The time trace of the local order parameters provides a useful qualitative indicator of chimera states, but determining the side length $2\delta+1$ of the local region surrounding each oscillator is non-trivial. As $\delta$ enlarges, more of the globally coherent domains are captured but these domains become merged with incoherent domains. In contrast, reducing $\delta$ will identify more incoherent and spatially localized domains but the global coherent domains will not be recognized. We thus decided that to characterize the chimera states, we should examine the coexistence of two spatially separated domains, in which one part of the oscillator network operates coherently while the other is incoherent. 
By using the time-variant topological features to track the time trace of these domains, we could better understand how chimera states are evolved, which would allow for a qualitative prediction of the chimera states in the early stage of the dynamics.

Here, we present a mapping to transform the $j$th oscillator to a point on a torus surface
\begin{align}
\varphi: \theta_j \to \left(x_{\theta_j}, y_{\theta_j}, z_{\theta_j}\right),
\end{align} 
where 
\begin{align}
 x_{\theta_j} &= (R_m+R_p\cos\theta_j)\cos\left(\dfrac{2\pi j}{N}\right),\\
 y_{\theta_j} &= (R_m+R_p\cos\theta_j)\sin\left(\dfrac{2\pi j}{N}\right),\\
 z_{\theta_j} &= R_p\sin\theta_j.
\end{align}
We set the major radius $R_m$ and minor radius $R_p$ of the torus as $R_m = 4$ and $R_p = 1$, respectively. Through this mapping, chimera states can be identified in higher-order topological structures such as loops. For example, in the globally synchronized state, the mapped points tend to distribute along one major loop on the torus surface, and more minor loops form as the number of incoherent regions increases.

\begin{figure*}
	\centering
	\includegraphics[width=17cm]{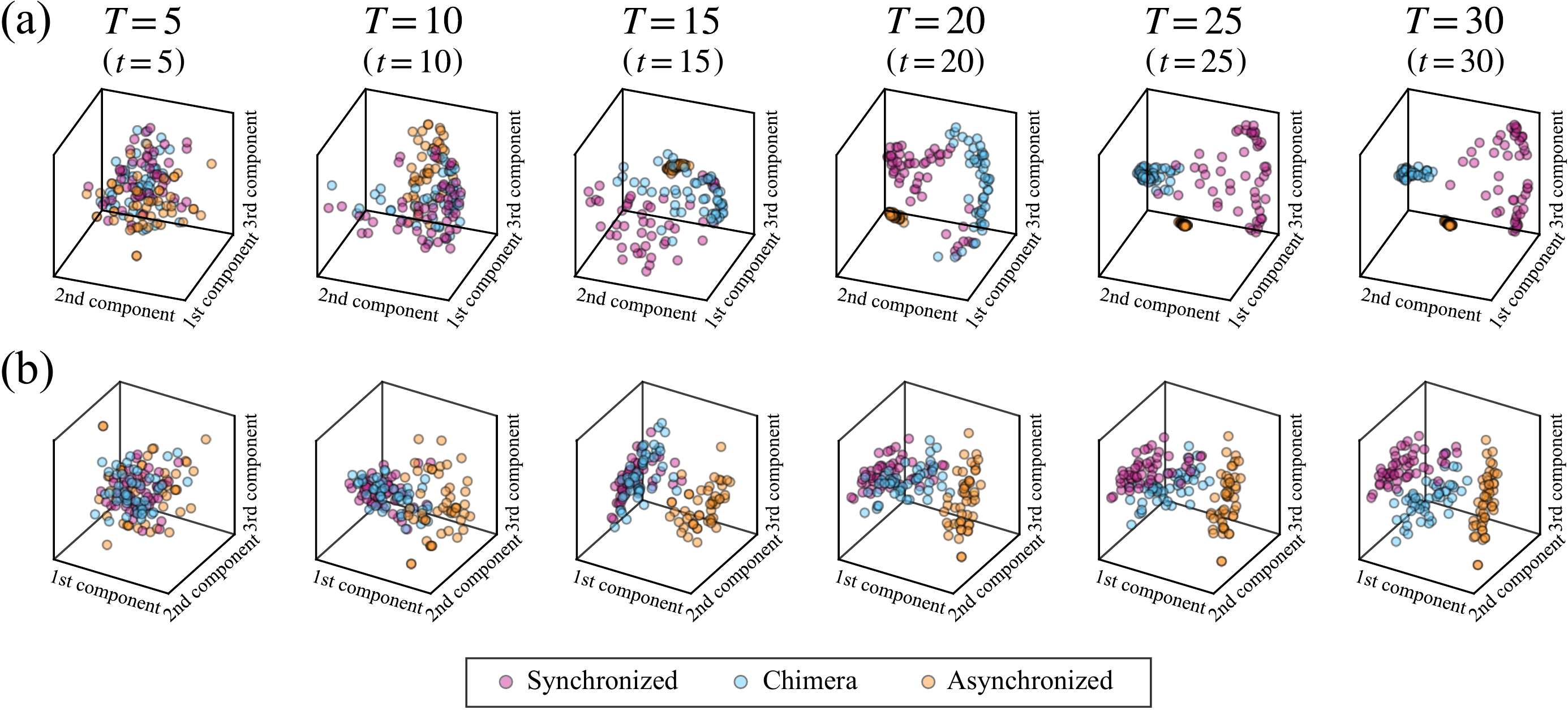}
	\caption{
	Dimensional reduction of the time-variant topological features of (a) the connected components (top row), and (b) loops (bottom row), obtained in the kernel principal component analysis. Shown are the distributions of each dynamical case at $T=5, 10, 15, 20, 25, 30$ (corresponding to $t=5, 10, 15, 20, 25, 30$) from left to right. Each point is in a synchronized (purple), chimera (blue), or asynchronized state (orange).}
	\label{fig:kpca_b}
\end{figure*}

Figures~\ref{fig:lop_pd}(b)(c) present the three-dimensional persistence diagrams of the loop patterns and the death-radius distributions of the connected components, respectively, during the phase evolution of the configurations corresponding to the examples in Fig.~\ref{fig:lop_pd}(a). 
To compute the three-dimensional persistence diagrams, we set the time step as $\tau_0=0, \tau_k-\tau_{k-1}=1$ $(k=1,2,\ldots,T-1)$.
The corresponding patterns of the mapped points in three-dimensional space at $t=40$ are illustrated in Fig.~\ref{fig:lop_pd}(d). 
At each time point, the point cloud formed one large loop around the torus hole.
This loop was born at a low radius and died at a high radius.
Over time, the loops corresponded to a thinning column in the three-dimensional persistence diagram [Fig.~\ref{fig:lop_pd}(b)].
The many loops with low birth and death radii presented as minor circles of point clouds on the torus surface.

In the global synchronized state, for example at $t \geq 25$ (top row of Fig.~\ref{fig:lop_pd}(a)), a single loop appeared along the torus tube in the mapped space, meaning that one point appeared in the three-dimensional diagram (Fig.~\ref{fig:lop_pd}(b)) at each $t$. 
The middle panel of Fig.~\ref{fig:lop_pd}(b) shows the emergence of coherent and incoherent dynamics (i.e., a chimera state). Small loops appeared around the minor circles of the torus surface, meaning that more points were generated along the time axis in the three-dimensional diagram of the loop patterns. The density of these points in the persistence diagram increased as the incoherent dynamics dominated the phase dynamics (Fig.~\ref{fig:lop_pd}(b), bottom panel). 
In the globally synchronized state, even when the phase values $\theta_i(t)$ were invariant with respect to $i$, they changed along time $t$. Therefore, the death radius of the large loop around the torus hole fluctuated over a large range.
This fluctuation was reduced when both coherent and incoherent dynamics emerged in the chimera state.
In the asynchronized state, the phase values $\theta_i(t)$ differed with both $i$ and $t$, so the death radius of the large loop did not greatly fluctuate. Therefore, a thinner column persisted in this case.

The differences among the oscillator phase dynamics were further evaluated by observing the death-radius distributions of the connected components (see Fig.~\ref{fig:lop_pd}(c)). In the globally synchronized state, connected components emerged and then quickly merged at almost the same death radii (Fig.~\ref{fig:lop_pd}(c), top panel). Conversely, in the asynchronized state, the mapped points of the oscillators on the torus surface were randomly distributed. The death radii were concentrated in the range 0-1, and their distributions were almost constant throughout the time evolution (Fig.~\ref{fig:lop_pd}(c), bottom panel). In the chimera state, the death radii of the connected components were smaller in the coherent region than in the incoherent region, and the death radius was widely distributed along the timeline (middle of Fig.~\ref{fig:lop_pd}(b)). It should be noted that time-variant topological features provide a novel means of recognizing chimera states in quantitative terms without having to rely on the tuning parameter $\delta$ of the local order parameter.

We now demonstrate that the kernel method based on the time-variant topological features can characterize the chimera states in early-stage dynamics. Specifically, we prepared 150 temporal phase data under different initial-phase conditions. The phase data were labeled as synchronized, chimera, or asynchronized (50 cases in each state). In general, global order parameters at the early stages cannot properly distinguish local structures such as chimera states. When labeling the dynamics in the present setting, we relied on the global order parameter $r(t)$ at sufficiently large $t$; specifically, we set synchronized states for $r(t=40) > 0.85$, chimera states for $0.45 \leq r(t=40) \leq 0.85$, and asynchronized states for $r(t=40) < 0.45$.
We did not quantitatively compare the topological features using the order parameters, but emphasized that the time-variant topological features can infer the long-time dynamic patterns at an early stage.

Figure~\ref{fig:kpca_b} displays the projection up to the third component of the kPCA of the kernel Gram matrix constructed from 150 three-dimensional persistence diagrams. 
To compute the three-dimensional persistence diagrams, we set the time step as $\tau_0=0, \tau_k-\tau_{k-1}=1$ $(k=1,2,\ldots,T-1)$. 
During the initial evolution ($T=5$), no clear differences appeared among the points representing different behaviors. 
The points representing the asynchronized states were clearly separated from the other states at $T=10$ and $15$ (i.e., $t=10, 15$), and all three states were separated around $T=20$ and $25$ (i.e., $t=20, 25$).
This separation also appeared in the three-dimensional diagrams of Fig.~\ref{fig:lop_pd}(b).

In the time trace of the local order parameters (Fig.~\ref{fig:lop_pd}(a)), the states became obviously separated at a relatively early time (around $t=10\text{--}15$). From this observation, one might question the advantage of the topological features over the local order parameters (the traditional indicators of chimera states).
Although Fig.~\ref{fig:lop_pd}(a) presented a specific example, the local order parameters well characterized the synchronization behaviors of oscillators.
However, their calculation requires careful calibration of $\delta$, which controls the parameter-computation region around each oscillator.
The same trade-off was clarified in the classification of multicluster synchronizations in Part A of this section (see Fig.~\ref{fig:gauss_acc}).
From this perspective, the time-variant topological features can obtain the global and robust features of oscillator behaviors independently of hyperparameters. That is, the separation between topological features can reveal novel behaviors of oscillator systems. In data analysis, this approach might infer or discover unknown properties of physical systems.

\section{Concluding remarks and discussion\label{sec:discussion}}
In this paper, we demonstrated that the time-variant topological features constructed from the phase evolution in oscillatory systems can be used to characterize the behavior of the dynamics, even in the early stages of the evolution. Such behaviors include global synchronization, multicluster synchronization, and chimera state emergence, which conventional global order parameters fail to sufficiently recognize. This indicates that our topological approach is an effective approach for understanding the phase dynamics of oscillators.

In previous applications of persistent homology to oscillatory systems, only the average temporal patterns were considered~\cite{stolz2017persistent, maletic2016persistent}. Our approach fundamentally differs from such an approach in that it allows us to trace the temporal patterns, which are more helpful to investigating the specific behavior of dynamics. Furthermore, by combining our approach with the machine learning kernel method, we provided an unsupervised scheme to characterize the phase dynamics without predefined label training. This aspect is highly significant from the physical perspective, since unknown dynamics can be revealed using this unsupervised scheme, including in terms of characterizing the different types of chimera state.

It remains unclear as to whether mapping from a set of oscillator phases to a point cloud can be regarded as optimal mapping. In fact, it can be argued that other mapping methods involving various manifolds could extract more meaningful and higher dimensional topological information. Moreover, in addition to the values of the phases, other information, such as the phase derivatives, could be used to construct the time-variant topological features. In view of this, we expect that our study will be successfully applied to more practical situations in the future, including research involving noisy environments, nonuniform coupling strengths, or asymmetrical network structures, all of which may have oscillator networks with topological configurations that change over time.

Networks of coupled oscillators represent many systems in neuroscience and biology. 
The functionality of these networks depends on the collective dynamics of the interacting oscillators. Understanding how the underlying properties of these networks (coupling topology, interaction strengths, and characteristics of the individual nodes) describe their collective dynamics is a challenging task. 
Our study focused on synchrony, the most typical collective behavior of oscillator networks.
We characterized the shape of the collective dynamics through the time-variant topological features computed in the simplest oscillator model, namely, the phase oscillator model, in which the dynamics are captured by the evolution of phase variables.
The idea is extendible to oscillatory networks beyond phase oscillators. For example, synchronous behaviors can be described in terms of frequency synchrony (coincidence of the oscillator frequencies). 
In these cases, the time-variant topological features can be calculated by mapping not the phase variables, but other descriptive variables of the oscillators.
One could then investigate how the emergence of topological features during the dynamics evolution relates to the functionality of oscillatory networks. 

\appendix
\begin{figure*}
    \centering
    \includegraphics[width=18cm]{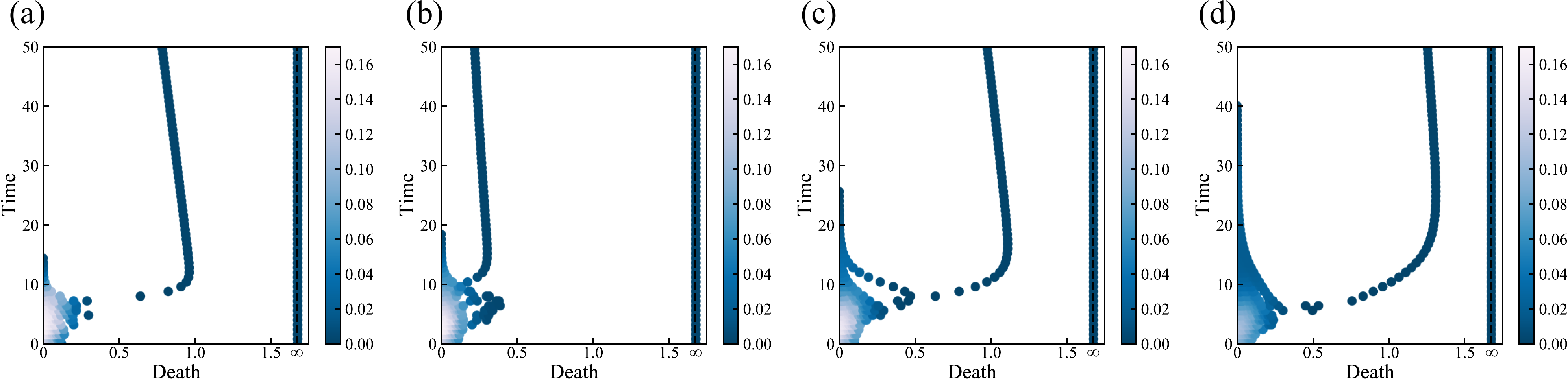}
    \caption{Examples of time-variant topological features corresponding to the networks with different structures: (a) 64-64, (b) 80-48, (c) 96-32, and (d) 112-16.
    Plotted are the death-radius distributions of the connected components during the phase evolutions of the oscillators.
    The colored bar to the right of each plot represents the probability densities of the death-radius distribution.
    }
\label{fig:asym}
\end{figure*}
\section{Fisher information metric}\label{appx:kernel}
The persistence Fisher kernel was constructed in Fisher information geometry, in which each $\rho_{\bD}$ in Eq.~\eqref{eqn:rho} is regarded as a point in the probability simplex $\bbP=\{\rho \mid \int_{\bOm} \rho(\bx)=1, \rho(\bx) \geq 0\}$.
To define the Fisher information metric between $\rho_i=\rho_{\bD_i}$ and $\rho_j=\rho_{\bD_j}$, we use the Hellinger mapping $h(\cdot) = \sqrt{\cdot}$, where the square root is an element-wise function that transform $\bbP$ into $\mathbb{S}_{+}=\{\chi | \int_{\bOm} \chi^2(\bx) = 1, \chi(x) \geq 0\}$. Here, $\mathbb{S}_{+}$ is the positive orthant of the sphere.
Therefore, we can naturally define the metric between $\rho_i$ and $\rho_j$ in $\bbP$ as the geodesic distance between $h(\rho_i)$ and $h(\rho_j)$ in $\mathbb{S}_{+}$.
This geodesic distance, known as the Fisher information metric~\cite{amari:2016:info}, is calculated as 
\begin{align}
d_\textup{F} &= \textup{arccos}\left( \langle h(\rho_i), h(\rho_j) \rangle \right) \\
 &= \textup{arccos}(\int_{\bOm} \sqrt{\rho_i(\bx)\rho_j(\bx)}d\bx),
\end{align}
where $\langle \cdot, \cdot \rangle$ is a dot product.

In practical implementations, the Fisher information metric between two diagrams $\bD_i$ and $\bD_j$
is computed over $\bOm=\bD_i\cup \bD_{j\Delta}\cup\bD_j\cup \bD_{i\Delta}$. We first write
\begin{align}
    \bar{\rho}_i = \rho_{\bD_i\cup \bD_{j\Delta}} = \left[ \dfrac{1}{Z}\sum_{\bp \in \bD_i \cup \bD_{j\Delta}}\cN(\bx;\bp, \nu\bI) \right]_{\bx \in \bOm},
\end{align}
where $\nu$ is the smoothing bandwidth and the normalization constant is given by
\begin{align}
Z = \sum_{\bx \in \bOm}\sum_{\bp \in \bD_i \cup \bD_{j\Delta}} \cN(\bx;\bp, \nu\bI).
\end{align}
We then compute $\bar{\rho}_j = \rho_{\bD_j\cup \bD_{i\Delta}}$ similarly as $\bar{\rho}_i$.
Finally, the Fisher information metric becomes 
\begin{align}
    d_\textup{F}(\bD_i, \bD_j) = \textup{arccos}\left( \langle h(\bar{\rho}_i), h(\bar{\rho}_j) \rangle \right).
\end{align}

\section{Asymmetric network structure}\label{appx:assym}

In this Appendix, we demonstrate that our time-variant topological features are applicable to asymmetric network structures, in which the modules are imbalanced.
We prepared four different network structures of $N=128$ oscillators. 
All oscillators were divided into an $n_1$-oscillator module and an $n_2$-oscillator module, where $n_1+n_2=N$. The networks were denoted by $n_1\text{--}n_2$; for example, the 80\text{--}48 structure means that 80 and 48 oscillators were assigned to the first and second modules, respectively. 
We generated networks with the following structures: 64\text{--}64, 80\text{--}48, 96\text{--}32, and 112\text{--}16. 
Note that the 80\text{--}48, 96\text{--}32, 112\text{--}16 networks were asymmetric whereas the 64\text{--}64 network was symmetric. In this task, we set $g_{ij}=2$ for oscillators in the same module and $g_{ij}=0.01$ for those in different modules.
The global coupling constant and natural frequency were $\kappa=1$ and $\omega_i=0$, respectively.

Figure~\ref{fig:asym} shows the death-radius distributions of the connected components during the phase evolutions of the oscillators.  At all ratios of $n_1$ and $n_2$, the number of connected components corresponded to the number of clusters in the oscillator dynamics. For example, the two connected components surviving from $t=40$ to $t=50$ indicate that at this stage of the evolution, the oscillators evolved into two clusters.
At the earliest time when only two clusters existed in the oscillator's dynamics, we observe that increasing the difference $|n_1-n_2|$ between the number of nodes in the two clusters increased this value  and lowered the density of components with low death radii.
As the modules were imbalanced, the oscillators in each module converged to clustered synchronization at different times.
Therefore, the time-variant topological features can reveal how network structures imply function through the synchronization behaviors of their oscillators.

\section{Classification methods}\label{appx:classify}

We constructed a vector of the temporal sequence $\br = \{r(\tau_0), r(\tau_1), \ldots, r(\tau_{T-1})\}$, where
$r(t)$ is the global order parameter defined in Eq.~\eqref{eqn:global:kuramoto}.
We also constructed vectors of the temporal sequences of the local order parameters defined by Eq.~\eqref{eqn:kura:local} for all coupling configurations.
The authors of ~\cite{arenas:2006:prl:synch} constructed the dynamic connectivity matrix among the oscillators from the local order parameters averaged over random initializations of the correlation between pairs of oscillators.
To classify the individual coupling configurations, we modified the dynamic connectivity matrix in Ref.~\cite{arenas:2006:prl:synch} as follows:
\begin{align}\label{eqn:dynamic:mat}
    \cD_\Theta(t)_{ij} = 
    \begin{cases}
    1, & q_{ij}(t)\geq\Theta \\
    0, & q_{ij}(t)<\Theta
    \end{cases},
\end{align}
where $\Theta$ is some threshold and the pairwise correlation is 
$
    q_{ij}(t) = \cos(\theta_i(t)-\theta_j(t)).
$
We constructed the global vectors describing the temporal evolution of the eigenvalue spectrum of $\cD_\Theta(t)$.
Among different trials of $\Theta$ ($0.9,0.8$, and $0.7$) $\Theta=0.99$  achieved the best result.

The input data were the constructed vectors $\bx_1, \bx_2, \ldots, \bx_M$ of coupling configurations. Each element in each vector was computed at each time point. 
In SVM, we therefore search for the hyperplanes separating the data belonging to different labels.
This kind of problem reduces to a dual constrained optimization problem in which the original data are represented by an $M\times M$ Gram matrix $K$ in the original space defined by $K_{ij}=\bx^\top_i\bx_j$.
If we classify the kernel of the persistence diagrams, the Gram matrix in the original space is replaced by the kernel Gram matrix of the diagrams, avoiding the conversion of each diagram into a separate vector. When solving the kernel Gram matrix, the SVM becomes a linear classifier  in the feature space. The implementation of SVM and its variants with kernel tricks are detailed elsewhere ~\cite{bishop:2006:prm,schlkopf:2018:kernel}.

\section{Effect of the time interval $\Delta \tau$}\label{appx:interval}

We also investigated the impact of time interval on the topological method that distinguishes different synchronization scenarios. Table~\ref{tab:table} lists the average test accuracies over 100 random train-test splits and their standard deviations at $t=8, 16$, and $32$ for different sampling intervals ($\Delta\tau=0.4, 0.8$, and $1.6$).
The final row of Table~\ref{tab:table} gives the result of extracting the topological features at a single point (the last time point).
As shown in this Table, lowering the time interval $\Delta\tau$ improved the classification accuracy. 
The topological features extracted at a single time point yielded the worst performance. Moreover, the effect of reducing the time interval became more vital at later times $t$.
These results demonstrate the importance of collecting the time-variant topological features over a wide temporal range.
However, computing the persistent homology and kernel at small sampling intervals over a wide time range is time- and memory-intensive.

\section{Effect of changing the number of selected oscillators}\label{appx:noscillators}
Finally, we examined the effect of randomly selecting a number of oscillators for the classification.
In this examination, we used the time-variant topological features at $\Delta\tau=0.8$ and $t=8, 16,$ and $32$, corresponding to $T=10, 20$, and $40$ in Fig.~\ref{fig:gauss_acc}, respectively.
Table \ref{tab:sample} shows the average test accuracies over 100 random train--test splits and their standard deviations after randomly selecting $100\%, 75\%, 50\%$, and $25\%$ of the oscillators.
As a reference for comparison, we computed the classification accuracies of the traditional methods using the global order parameters and dynamic connectivity matrix. In both comparative methods, we included 100\% of the oscillators in each sample.

As clarified in Table~\ref{tab:sample}, decreasing the number of oscillators decreased the classification accuracy; nonetheless, our approach based on the time-variant topological features outperformed the other methods, even when the number of oscillators was reduced to 75\%.
This result demonstrates the strong effectiveness of the topological features. The time-variant topological features obtain the essential behaviors even when the oscillators are sparsely configured. In contrast, the utility of the global information (global order parameter and dynamic connectivity) is limited when the configuration is sparse.

\begin{table}[htb]
  \caption{Averages and standard deviations  (mean$\pm$sd) of the classification accuracies (\%) of synchronized states using the time-variant topological features computed from the connected components at $t=8, 16,$ and $32$ with different time intervals ($\Delta\tau=0.4, 0.8$, and $1.6$) and (last row) using the topological features at the last time point}
    \centering
    \begin{tabular}{cccc}
        \hline\hline
        $\Delta\tau$ & $t=8$ & $t=16$ & $t=32$ \\ \hline
        0.4 & $68.7\pm4.0$ & $73.7\pm3.6$ & $80.1\pm2.5$\\
        0.8 & $68.4\pm4.1$ & $70.0\pm4.9$ & $76.9\pm3.4$\\
        1.6 & $67.7\pm4.4$ & $70.9\pm4.4$ & $73.6\pm4.3$\\ \hline
        Single & $63.9\pm3.8$ & $64.9\pm2.9$ & $64.4\pm2.9$ \\ \hline\hline
    \end{tabular}
    \label{tab:table}
\end{table}

\begin{table}[htb]
    \caption{
    Averages and standard deviations (means $\pm$sd) of the classification accuracies (\%) of synchronized states based on the time-variant topological features extracted from the connected components of 100\%, 75\%, 50\%, and 25\% randomly selected oscillators, and from the global order parameter and dynamic connectivity matrix of all oscillators (last two rows).
    }
    \centering
    \begin{tabular}{cccc}
        \hline\hline
        Method & $t=8$ & $t=16$ & $t=32$ \\ \hline
        100\% oscillators & $68.4\pm4.1$ & $70.0\pm4.9$ & $76.9\pm3.4$ \\
        75\% oscillators & $67.3\pm4.5$ & $67.5\pm4.5$ & $72.9\pm4.2$ \\
        50\% oscillators & $61.7\pm3.4$ & $65.2\pm3.4$ & $69.4\pm4.4$ \\
        25\% oscillators & $56.4\pm3.1$ & $63.1\pm3.2$ & $64.3\pm3.6$ \\
        \hline
        Global order parameter & $51.7\pm4.4$ & $62.4\pm3.8$ & $71.9\pm3.1$ \\
        Dynamic connectivity  & $59.3\pm3.7$ & $63.5\pm3.6$ & $67.2\pm3.3$ \\
\hline\hline
    \end{tabular}
    \label{tab:sample}
\end{table}

\end{document}